\documentclass{ctr_summer}

\usepackage{ctrfont}
\usepackage{natbib}
\usepackage{caption}
\usepackage{subfig} %
\usepackage{cleveref}
\usepackage{csquotes}
\usepackage[dvips]{graphicx}
\usepackage{psfrag}
\usepackage{epsfig}
\usepackage{amsmath,rotating}

\graphicspath{{figures/}}

\usepackage{url}

\usepackage{color}

\title{Correction scheme for point-particle two-way coupling applied to nonlinear drag law}

\shorttitle{Correction scheme for point-particle two-way coupling applied to nonlinear drag law}

\author[J.A.K. Horwitz, A. Mani]{J.A.K. Horwitz\thanks{ Email address for correspondence: horwitz1@stanford.edu}, A. Mani\\Department of Mechanical Engineering\\ Stanford University}

\shortauthor{Horwitz and Mani}

\begin{document}
\setcounter{page}{1}

\maketitle
Drag laws for particles in fluids are often expressed in terms of the undisturbed fluid velocity, defined as the fluid velocity a particle sees before the disturbance develops in the fluid.  In two-way coupled point-particle simulations the information from the undisturbed state is not available and must be approximated using the disturbed velocity field. \citet{horwitz-2016} recently developed a procedure to estimate the undisturbed velocity for particles moving at low Reynolds number and obeying the Stokes drag law. Using our correction, we demonstrated convergence of numerical simulations to expected physical behavior for a range of canonical settings. In this paper we further extend and examine that correction scheme for particles moving at finite Reynolds number, by considering the Schiller-Nauman drag law. Tests of particle settling in an otherwise quiescent fluid show reasonable predictions of settling velocity history for particle Reynolds numbers up to ten. The correction scheme is shown to lead to $\geq 70\%$ reduction in particle velocity errors, compared to standard trilinear interpolation of disturbed velocities.  We also discuss the modelling of unsteady effects and their relation to Stokes number and density ratio. Finally, we propose a regime diagram to guide scheme selection for point-particle modelling.
\\
\hrule

\section{Introduction}
Many natural and industrial processes comprise the transport of small solid particles by a gas. Sand storms can pose health risks for Earth communities as well as challenges to sustainable human habitat on Mars \citep{Kok-2012}. The ash ejected during volcanic eruptions \citep{Lavallee-2015} can be hazardous to air traffic. In addition, small particulate matter $O(\mu m)$ created during the combustion of gasoline and coal has been linked to increases in daily mortality \citep{Laden-2000}. Another application concerns solar receiver technologies that utilize absorbing particles; one such design is under investigation by the PSAAP2 program \citep{Pouransari-2017, Farbar-2016}. A channel flow of air is to be seeded with $O(10 \mu m)$ nickel particles and irradiated. The goal is to increase the outlet gas temperature by having particles act as intermediaries, absorbing the solar radiation and quickly convecting their internal energy volumetrically to the surrounding translucent air. \par
A commonly used tool to study particle-laden flows is via Euler-Lagrange numerical simulation. In this methodology, the carrier fluid phase is simulated in a static frame while each piece of dispersed phase material is tracked in a mesh-free approach. The fluid equations comprise the Navier-Stokes equations augmented by appropriate body forces (e.g. gravity) and particle forces. Particles obey Newton's second law and experience a change in their velocity owing to the hydrodynamic interactions with the fluid. However, resolving the detailed interaction around each particle for most applications is prohibitively expensive so that the point-particle assumption is often adopted. Under this assumption, instead of calculating the coupling force by integrating the fluid stress over the surface of a particle, a force model is specified by the user owing to some intuition regarding the physics of the problem. \par
\begin{figure}
  \centering
  \subfloat  {\includegraphics[trim={2.5cm 0.5cm 3.5cm 0.5cm},clip, width=105mm]{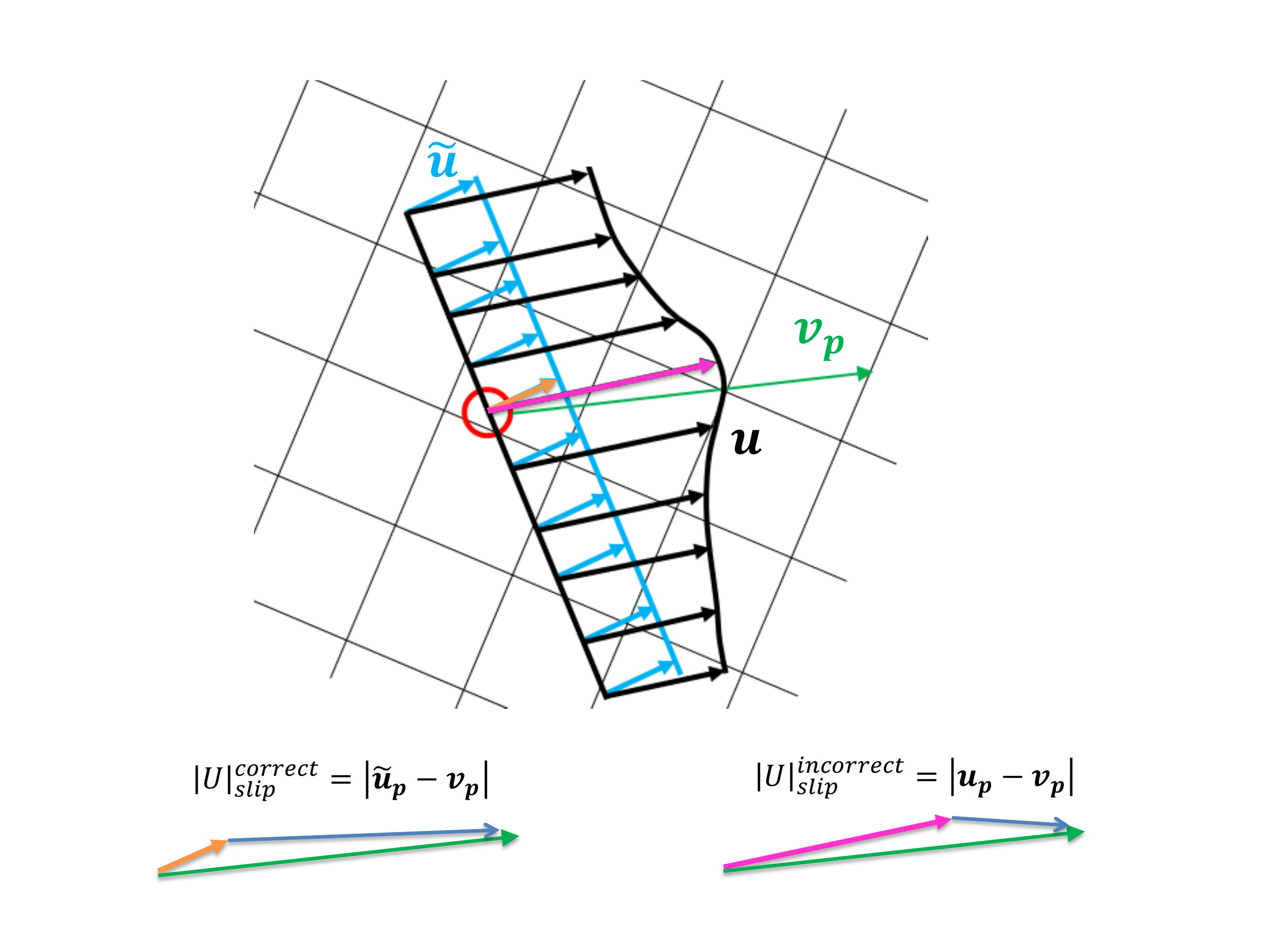} \label{disturbB}}
  \caption{Comparison of different velocities in Euler-Lagrange simulation of two-way coupled flow: $\boldsymbol{v_p}$ represents the particle velocity, $\boldsymbol{u}$ is the fluid velocity field, and $\boldsymbol{\tilde{u}}$ is the undisturbed velocity field (color online).}
  \label{fig:fig1}
\end{figure}

When the mass loading of particles is small, it is common to approximate the system as one-way coupled. In this limit, while particles experience resistance owing to the fluid, the fluid does not feel the presence of particles. However, when the mass loading of particles becomes $O(1)$, then the particles can play a substantial role in modifying fluid statistics, and reaction of the drag on the fluid phase must be explicitly accounted for.
Recently, the effect of two-way coupling on particle dynamics has been the subject of several works \citep{subramaniam_et_al_2014}, \citep{gualtieri-2015}, \citep{ireland-2016}, and \citep{horwitz-2016}. These studies were explored to remedy the observation by \citet{boivin-1998} that two-way coupled point-particles create disturbance flows which contaminate the fluid velocity in the near-field of the particles. This creates a numerical problem because calculation of the drag force often requires knowledge of the fluid velocity in the absence of the disturbance introduced by the presence of the particle. 
This idea is illustrated in Figure~\ref{fig:fig1}. Standard drag laws are a function of the difference between the particle's velocity $\boldsymbol{v_p}$ and the undisturbed fluid velocity at the particle location $\boldsymbol{\tilde{u}_p}$. While the particle velocity is readily obtained from the Lagrangian particle equations, the undisturbed flow is not directly accessible. When particles are two-way coupled to the fluid, the drag force experienced by the particle is fed back to the fluid. This creates a disturbance flow in the fluid which means the fluid velocity interpolated to the particle location, $\boldsymbol{u_p}$, is not equal to the undisturbed fluid velocity, $\boldsymbol{\tilde{u}_p}$. As the disturbance flow develops in the fluid, the difference between disturbed, $\boldsymbol{u_p}$, and undisturbed, $\boldsymbol{\tilde{u}_p}$, fluid velocity grows meaning that numerical implementations relying on the disturbed fluid velocity will be in error. \cite{horwitz-2016} showed this error was of the order of the particle size relative to the grid spacing, $\Lambda = d_p/dx$.

Under the assumption that particles obey Stokes drag, $\boldsymbol{F_d} = 3\pi\mu d_p(\boldsymbol{\tilde{u}_p}-\boldsymbol{v_p})$, accurate predictions for particle settling velocity were found when the undisturbed fluid velocity was well-predicted. The methods developed in \citet{gualtieri-2015}, and \citet{ireland-2016} are based on analytical solutions to a regularized point-force while the scheme developed in \citet{horwitz-2016} also includes specific effects of numerical discretization in their correction scheme. \par 
 Having developed accurate schemes to estimate the undisturbed fluid velocity for two-way coupled Stokesian point-particles is an important step towards validation of point-particle simulations in more complicated regimes with higher fidelity approaches (fully resolved particles and experiments.) However, the assumption that Stokes drag is the coupling force confines the validity of these schemes to low particle Reynolds numbers, which is a limiting condition considering practical applications. To extend the utility of two-way coupled point-particle methods, it is necessary to verify such methods outside the Stokes limit. \par
In this work, we apply the correction scheme of \cite{horwitz-2016} developed for particles obeying the Stokes drag law to particles settling at higher Reynolds numbers in an otherwise motionless fluid. We assume the particles obey the Schiller-Nauman correlation, $\boldsymbol{F_d} = \boldsymbol{F_{Stokes}} \cdot(1+0.15Re_p^{0.687} ) $ \citep{clift-1978}. We explore two primary questions: is the scheme developed in \citet{horwitz-2016} to compute $\boldsymbol{\tilde{u}_p}$ appropriate when the drag law is not Stokesian, and if so, over what range of Reynolds numbers can that correction scheme be used? Though the correction scheme in \citet{horwitz-2016} assumes certain Stokesian symmetries in the near-field of the particle, it seems the deviation from this ideal scenario changes slowly with Reynolds number. This is explained qualitatively by the analytical solution to the Navier-Stokes equations subject to a point force \citep{batchelor-1967}. 

\section{Methods}
The fluid and particle solvers are based on the code developed by \citet{pouransari-2015}. The Navier-Stokes equations augmented by a numerically regularized point-force are solved using 2nd order finite differences on a staggered grid. The resulting Poisson equation to enforce incompressiblity is solved directly with fast Fourier transforms. Particle positions and velocities are tracked in their respective Lagrangian frames. Both fluid and particle equations are updated with explicit 4th order Runge-Kutta time stepping. Particle momentum is coupled to fluid momentum via the drag force that occurs in their respective dynamic equations. Once the drag force model is chosen, the algorithm is closed by computing the undisturbed fluid velocity, $\boldsymbol{\tilde{u}}$, and projecting the Lagrangian force to the fixed Eulerian grid. The undisturbed fluid velocity is estimated by the formula $\boldsymbol{\tilde{u}}=\boldsymbol{u} +Cdx^2\nabla^2\boldsymbol{u}$, where $\boldsymbol{u}$ is the interpolated fluid velocity at the location of the particle and $C$ is a correction coefficient which depends on the non-dimensional particle diameter to grid spacing, $\Lambda = d_p/dx$ and the position of the particle within the cell. The $C$ coefficients can be found in \citet{horwitz-2016}. Once the drag force is calculated using the undisturbed velocity, the force is projected to the fluid grid using reverse trilinear interpolation. For further details on the point-particle algorithm, the reader is referred to the extensive discussion  found in \citet{horwitz-2016}.

\subsection{Specification of coupling force}
In this study, we release a point-particle from rest in a viscous fluid and allow it to settle under gravity. We assume the particle obeys the Schiller-Nauman drag correlation. This drag law takes the form: 
\begin{equation}\label{eq:1}
\boldsymbol{F_d}=\frac{m_p}{\tau_p}(\boldsymbol{\tilde{u}_p} -\boldsymbol{v_p} )∙(1+0.15Re_p^{0.687} )
\end{equation}

Here, $Re_p=|(\boldsymbol{\tilde{u}_p}-\boldsymbol{v_p} ) d_p/\nu|$, where  $\boldsymbol{\tilde{u}_p}$, $\boldsymbol{v_p}$, $d_p$, and $\nu$ are respectively the undisturbed fluid velocity evaluated at the particle location, the particle velocity, diameter, and fluid kinematic viscosity. In \eqref{eq:1}, $m_p$ and $\tau_p$ are respectively the particle mass and inertial relaxation time. Newton's 2nd law gives the particle equation in the Lagrangian frame:

\begin{equation}\label{eq:2}
\frac{d\boldsymbol{v_p}}{dt}=  \frac{\boldsymbol{\tilde{u}_p}-\boldsymbol{v_p} }{\tau_p}  (1+0.15Re_p^{0.687} )+\boldsymbol{g}(1-\rho_f/\rho_p )
\end{equation}

In \eqref{eq:2}, $\boldsymbol{g}$ is gravity, and $\rho_f$ and $\rho_p$ are respectively the fluid and particle density. The steady-state settling velocity (with $\boldsymbol{\tilde{u}} = \boldsymbol{0}$) satisfies:

\begin{equation}\label{eq:3}
u_s\bigg[1+0.15\Big(\frac{u_s d_p }{\nu}\Big)^{0.687}\bigg] =g\tau_p(1-\rho_f/\rho_p )
\end{equation}
Two methods of evaluating $\boldsymbol{\tilde{u}_p}$ are compared. The first method assumes that $\tilde{u}_p$ is equal to the disturbed fluid velocity evaluated at the particle location. This velocity is calculated using trilinear interpolation, which was found to be the most accurate method among conventional interpolation schemes in the absence of accounting for the undisturbed velocity \citep{horwitz-2016b}. The second method uses the procedure described in \citet{horwitz-2016} to estimate the undisturbed fluid velocity (zero for this flow) from discretely available velocity points in the neighborhood of the particle. No change is made to the $C$ coefficients reported in that work. Because the fluid-particle system is treated as two-way coupled, the negative of equation \eqref{eq:1}, normalized by the local fluid volume, is applied to the right hand side of the fluid momentum equation as a force density.  The disturbance velocity field created by the force density is what gives rise to the difference between the computed (disturbed velocity field), and the undisturbed velocity field in the absence of the particle.


We compare the Schiller-Nauman settling histories to the settling histories assuming the particles obey Stokes drag. All calculations are performed on a $128^3$ grid which was found to be sufficient for this type of verification problem \citep{horwitz-2016}.

\subsection{Dimensionless Parameters}
Similar to the previous work, we vary three parameters, the terminal particle Reynolds number $Re_p$, the non-dimensional particle size $\Lambda$, and the particle Stokes number, $St_\Delta=\tau_p/\tau_{visc}$. Because there is no imposed undisturbed flow in this problem, only the particle Reynolds number is a free physical parameter. However, two-way coupled point-particle simulation necessarily introduces numerical parameters owing to the fact that a singular source term is being numerically regularized. In \cite{horwitz-2016}, it was demonstrated that prediction of particle settling velocity in the absence of estimating the undisturbed fluid velocity depends strongly on $\Lambda$. 

The second numerical parameter (which can become a physical parameter in unsteady or turbulent flows) is the Stokes number, defined as the ratio of the particle relaxation time $\tau_p$ to the resolved viscous time of the grid, $\tau_{visc}=dx^2/\nu$. In direct numerical simulation, the smallest fluid scales in turbulence are the same order as the grid spacing, so that this definition of the Stokes number carries the same amount of information as a Kolmogorov based Stokes number \citep{pope-2000}. The reason for exploring a Stokes number dependence in this study is that while the Stokes number in the physical problem only controls the non-dimensional time at which a particle assumes some fraction of its terminal velocity, this parameter also represents numerically the transit time of the particle compared with the diffusion time of the disturbance field created by the particle. Only when $St_\Delta$ is large can one assume the disturbance flow establishes quickly and is quasi-steady, allowing the time-static correction scheme of \citet{horwitz-2016} to be justified. Conversely, when the Stokes number is order unity or less, the transit of the particle may couple to the disturbance flow in an unsteady way. In this study, we assume a steady drag correlation (which is a common assumption in the literature), and it is therefore important to test in what regimes the physical assumption of quasi-steadiness is consistent with the results of numerical calculations employing this assumption. 

\section{Results and Discussion}
In this study we examine four particle Reynolds numbers, $Re_p = [0.1, 1, 5, 10]$, three Stokes numbers, $St_\Delta = [0.5,5,25]$, and three non-dimensional particle sizes $\Lambda = [0.1,0.5,1]$. The results of the Reynolds number study are shown in Figure $2$. Shown are the numerical predictions for correction and trilinear schemes applied to a two-way coupled point-particle obeying the Schiller-Nauman correlation. Schiller-Nauman settling velocity histories based on numerical solutions of \eqref{eq:2} (with $\boldsymbol{\tilde{u}} = \boldsymbol{0}$) along with analytical Stokes settling solutions are shown for comparison. All results are normalized by the Schiller-Nauman steady-state settling velocity. For all of these cases, the correction scheme is superior to trilinear interpolation. While the correction coefficients presented in \citet{horwitz-2016} were developed for the zero Reynolds number limit, we observe that the same coefficients offer reasonable prediction of the undisturbed fluid velocity up to $Re_p$ of about ten. The maximum steady-state error for the correction scheme is about $9\%$ for $Re_p = 5$. 


\begin{figure}
  \centering
 . 
  \subfloat[$Re_p = 0.1$] {\includegraphics[trim={3.5cm 8.0cm 3.5cm 8.5cm},clip, width=65mm]{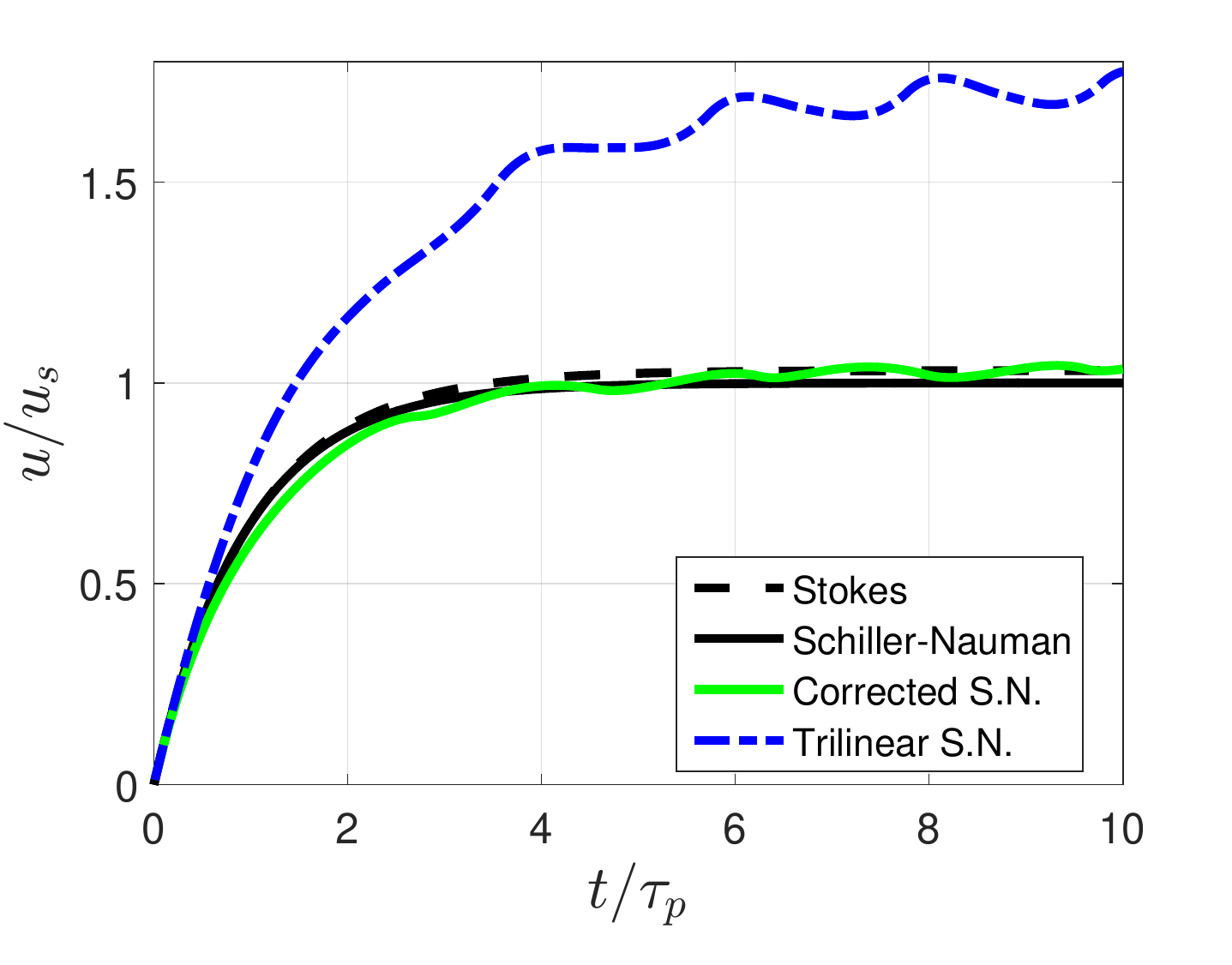} \label{fig:L1_R01_St5}}
  \subfloat [$Re_p = 1$] {\includegraphics[trim={3.5cm 8.0cm 3.5cm 8.5cm},clip, width=65mm]{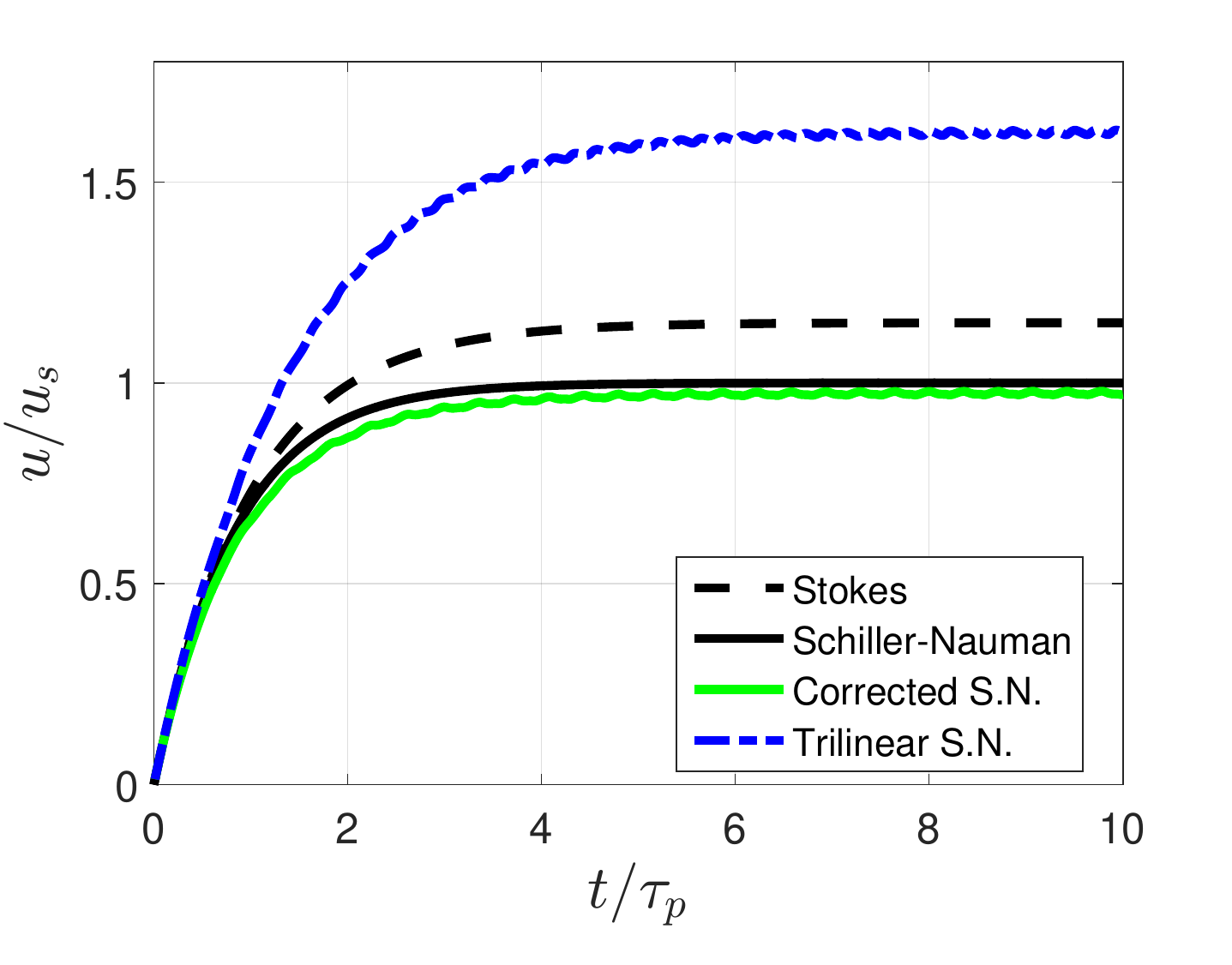} \label{fig:L1_R1_St5}}
  
  \subfloat[$Re_p = 5$] {\includegraphics[trim={3.5cm 8.0cm 3.5cm 8.5cm},clip, width=65mm]{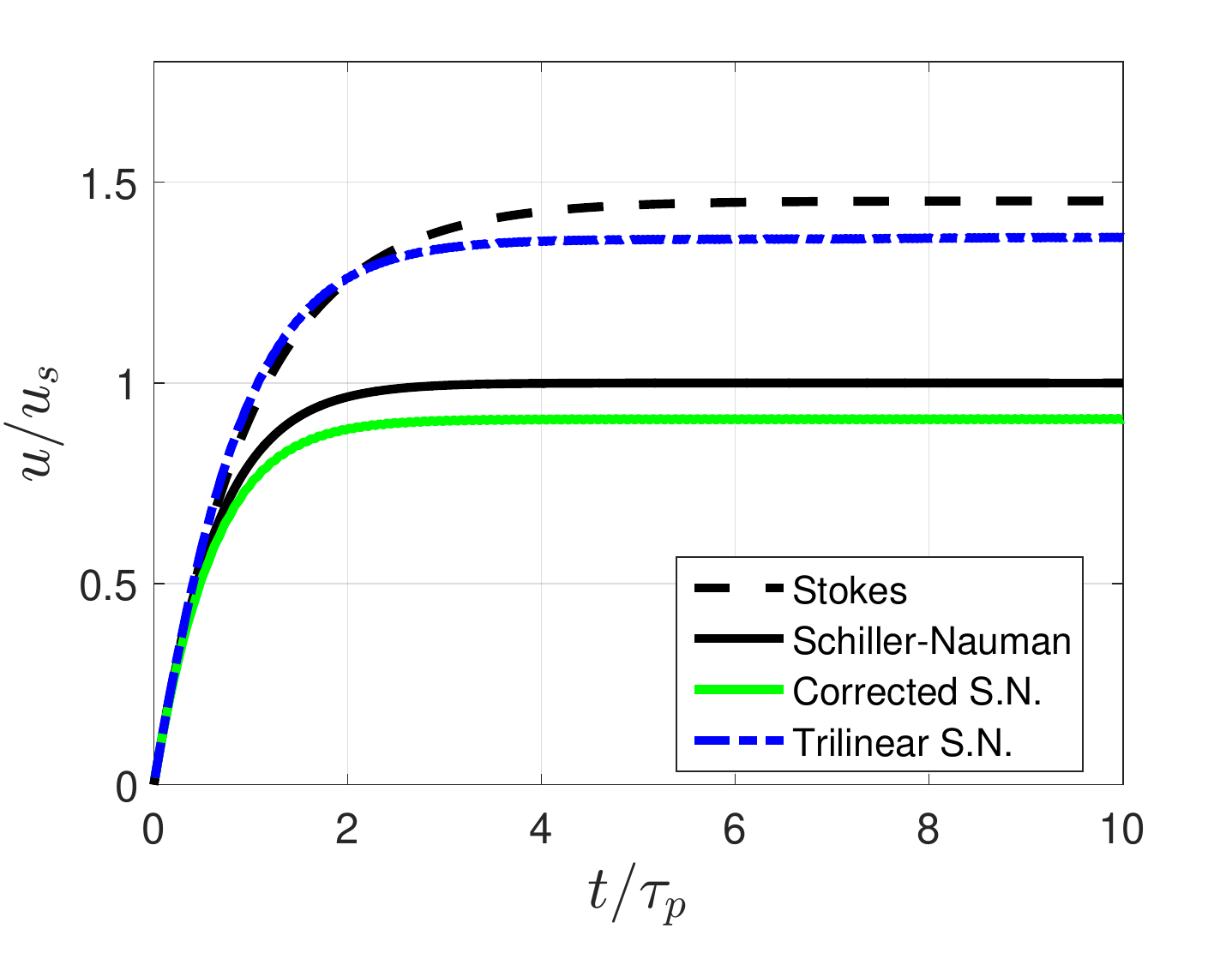} \label{fig:L1_R5_St5}}
  \subfloat[$Re_p = 10$] {\includegraphics[trim={3.5cm 8.0cm 3.5cm 8.5cm},clip, width=65mm]{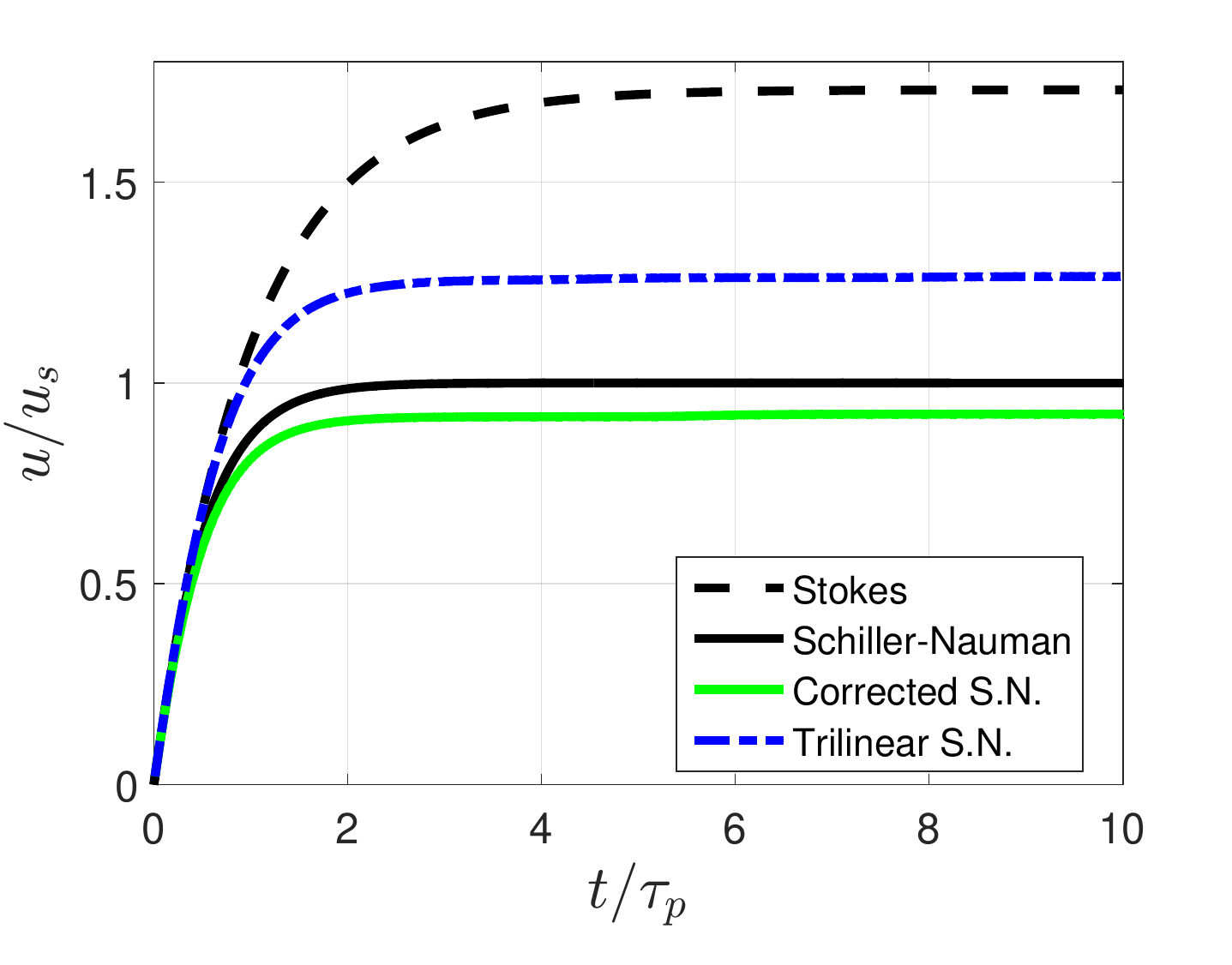} \label{fig:L1_R10_St5}}
  \caption{Settling velocity histories for different terminal Reynolds numbers, $\Lambda = 1$, $St_\Delta = 5$ (color online).}
  \label{fig:fig2} %
\end{figure}


\begin{figure}
  \centering
    \subfloat[$\psi/\nu$]{\includegraphics[trim={3.0cm 8.5cm 3.0cm 8.5cm},clip, width=65mm]{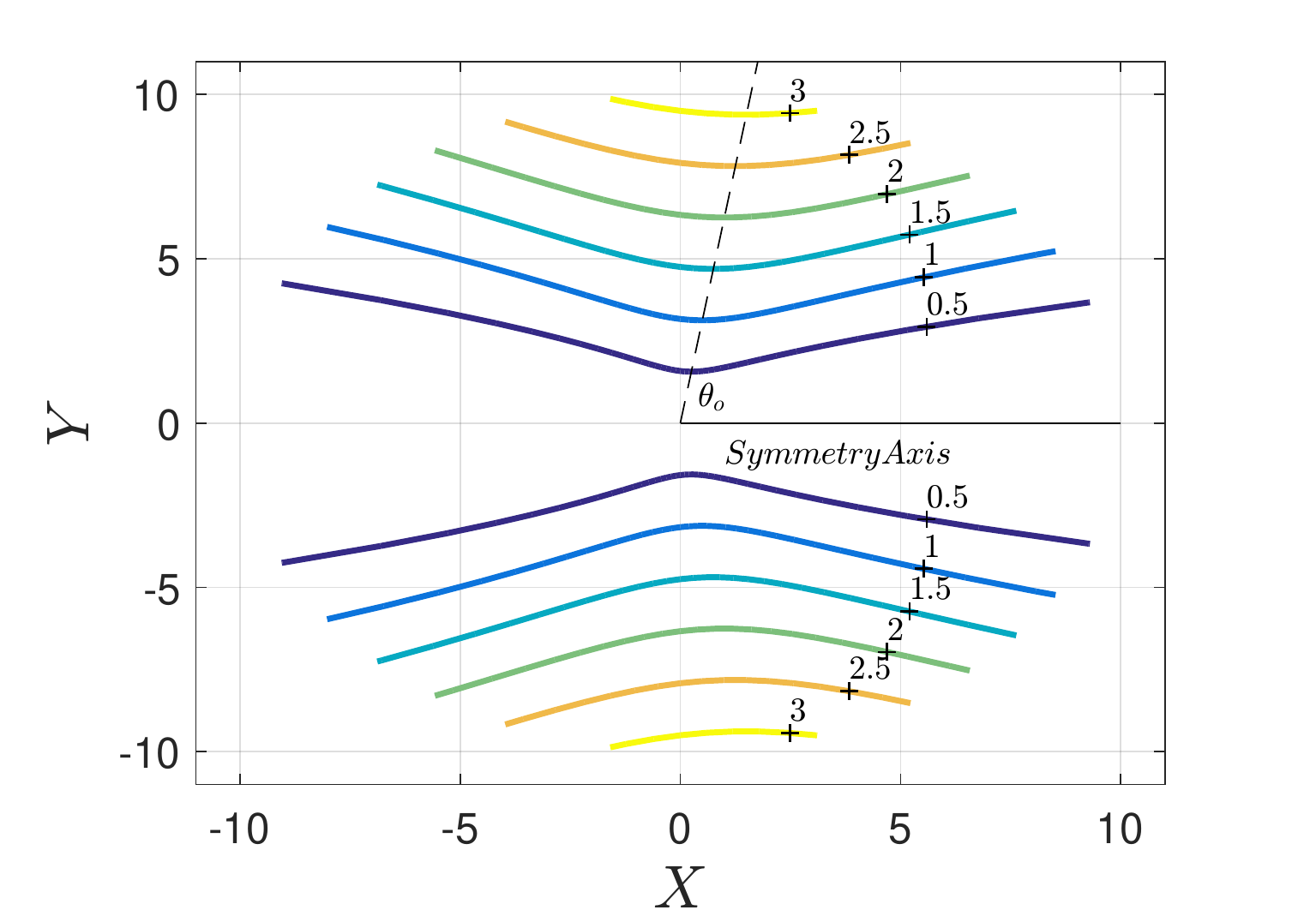} \label{fig:streamfunction}}
    \subfloat[Stream function angle]{\includegraphics[trim={3.0cm 8.5cm 3.0cm 8.5cm},clip, width=65mm]{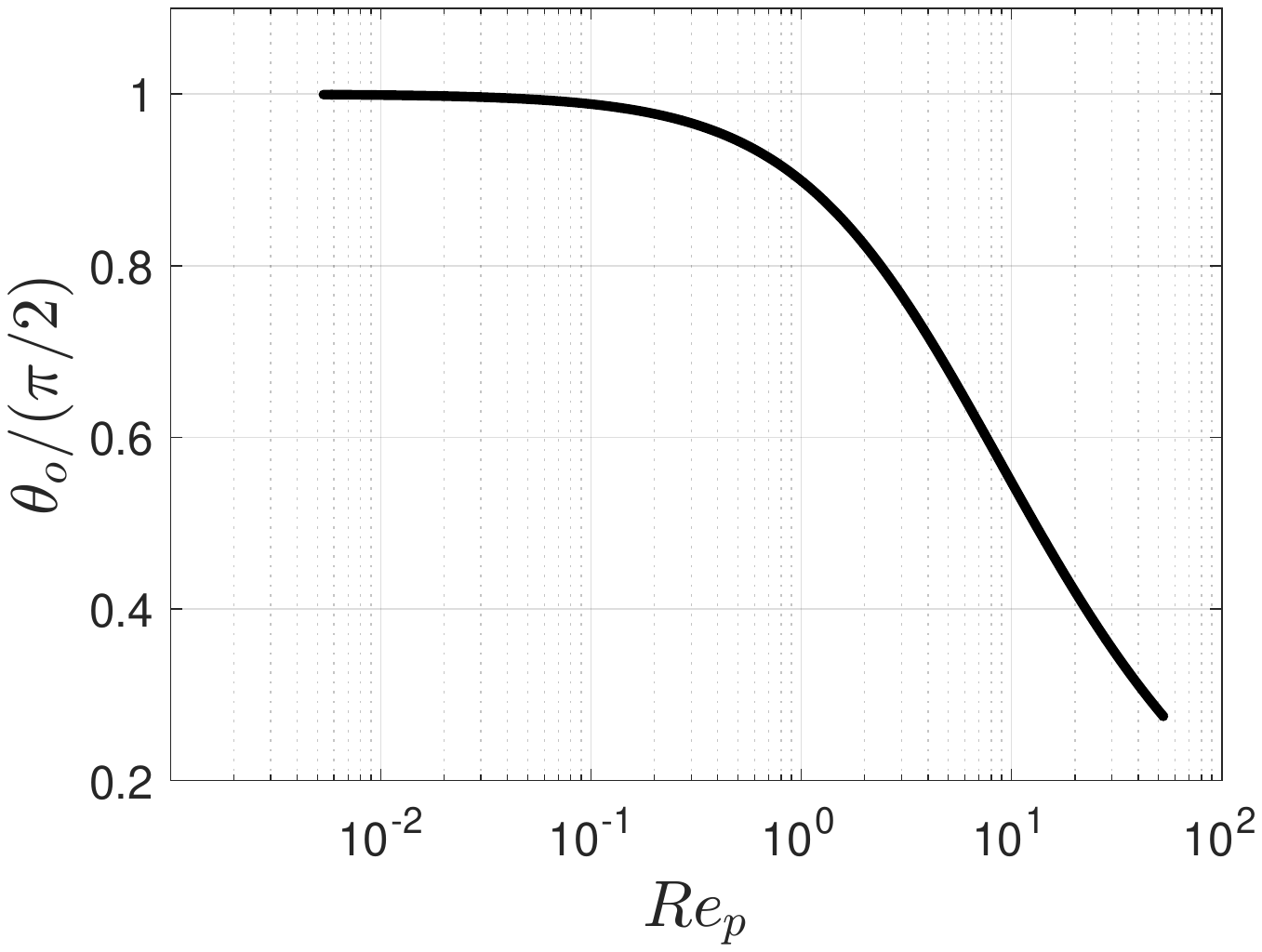} \label{fig:angle}}
  \caption{(a) Contours of stream function for point-force solution \citep{batchelor-1967} at $Re_p = 1$, and (b) Normalized stream function angle vs. Reynolds number show qualitative variation of a fluid's response to a point force (color online).}
  \label{fig:fig3}
\end{figure}

We may ask why the correction scheme developed in \cite{horwitz-2016} is able to reasonably predict particle settling outside the Stokesian regime. The point-particle equations are simply the Navier-Stokes equations augmented with a numerically regularized point force. As the point-particle moves, so too does the point force. A related problem is the steady jet flow which is a solution to the full steady and axisymmetric Navier-Stokes equations subject to a stationary point force \citep{batchelor-1967}. The solution depends on a free parameter $\alpha$ which, though independent of dimensional variables, can be defined in such a way as to predict the analytical force felt by a stationary sphere of diameter $d_p$ held fixed in a uniform flow of velocity $U$ in a fluid whose kinematic viscosity is $\nu$. In the low Reynolds number limit, a consistent expression for $\alpha$ is $\alpha=(16/3)Re_p^{-1}$. In other words, this definition of $\alpha$ implies $F = 3\pi\mu d_p U$ as $Re_p \rightarrow 0$, where $Re_p = Ud_p/\nu$. The salient observations from this solution are firstly, in the high $\alpha$ (low $Re_p$ limit), the stream function at large distances is Stokesian, independent of the nature of how the force is applied, and dependent only on the force magnitude. In other words, the streamlines far away from a point force are identical to those which would be created by a uniform flow over a finite sized particle in the low Reynolds number limit. Secondly, the deviation from Stokesian flow can be qualitatively measured by the nature of the streamlines. An angle can be constructed between the point at which streamlines are a minimum distance from the axis of symmetry, as in Figure~\ref{fig:fig3} (a). The formula is: $cos\theta_o = (1+\alpha)^{-1}$. In the limit $\alpha\gg1,$ ($Re_p\ll1 $), the streamlines are symmetric, so that $\theta_o=\pi/2, (〖90^{o})$. The normalized minimum stream function angle vs. Reynolds number is shown in Figure~\ref{fig:fig3} (b). We can see that the stream function angle is a smoothly varying function of the Reynolds number. At unity Reynolds number, the variation in the stream function angle is about $9^o$ with respect to the Stokes regime. The correction scheme evidently matches the analytical solution with low error at unity Reynolds number because the discretized equations have a disturbance flow which depends (in the near field of the particle) on the support of the projection stencil for the drag force. This causes the numerical solution to be regular and deviate from the continuous solution  \citep{batchelor-1967} in the near field of the particle. Therefore, while the correction scheme presented here assumes Stokesian symmetries, the regularity of the source term in our simulations reduces the deviation from Stokesian symmetry allowing satisfactory predictions of settling velocity even at a relatively high Reynolds number, $Re_p = 10$. 

\begin{figure}
  \centering
  \subfloat[$Re_p = 0.1$] {\includegraphics[trim={3.5cm 8.5cm 3.5cm 8.5cm},clip, width=65mm]{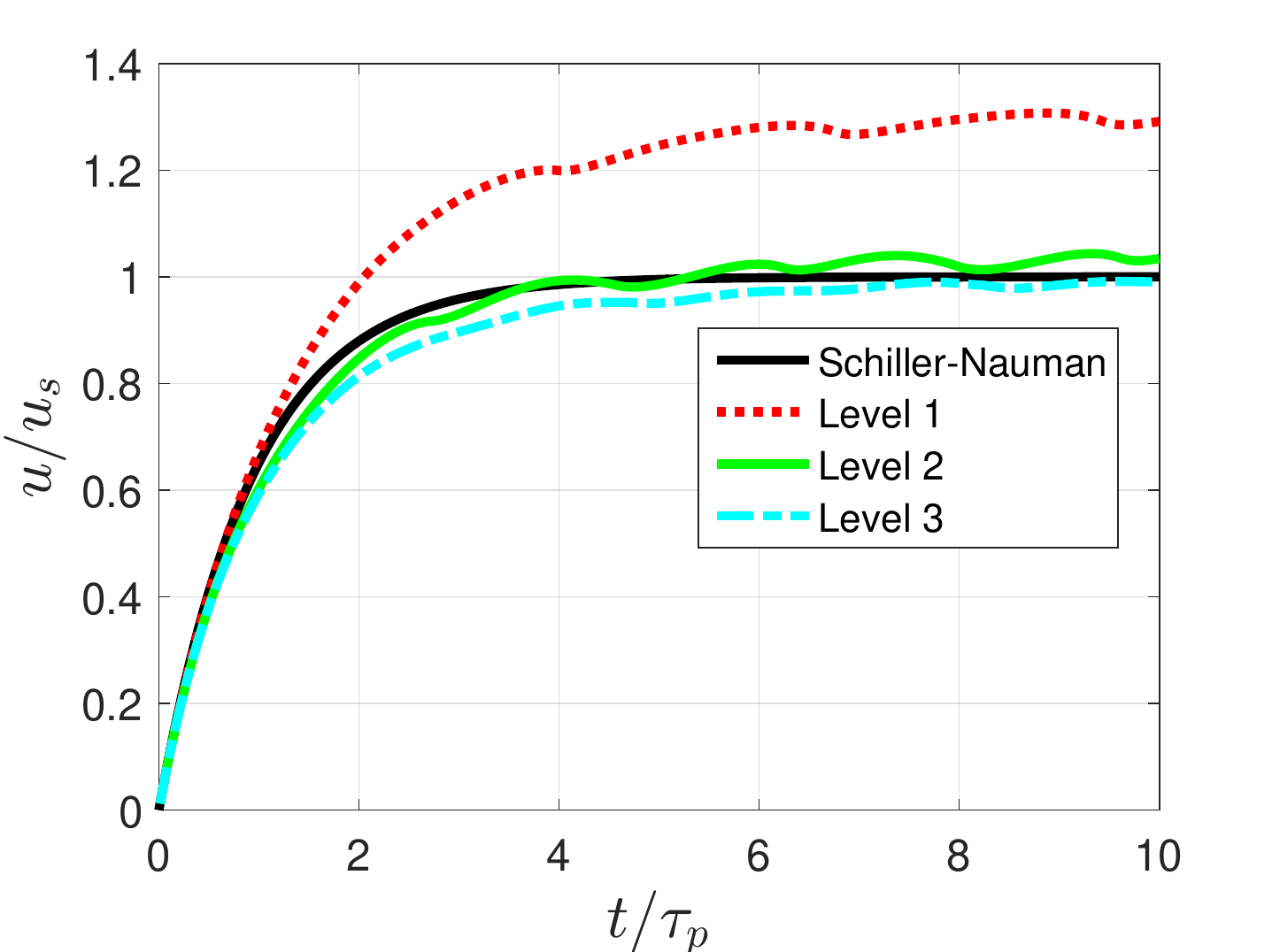} \label{fig:ref_L1_R01_St5}}
  \subfloat [$Re_p = 10$] {\includegraphics[trim={3.5cm 8.5cm 3.5cm 8.5cm},clip, width=65mm]{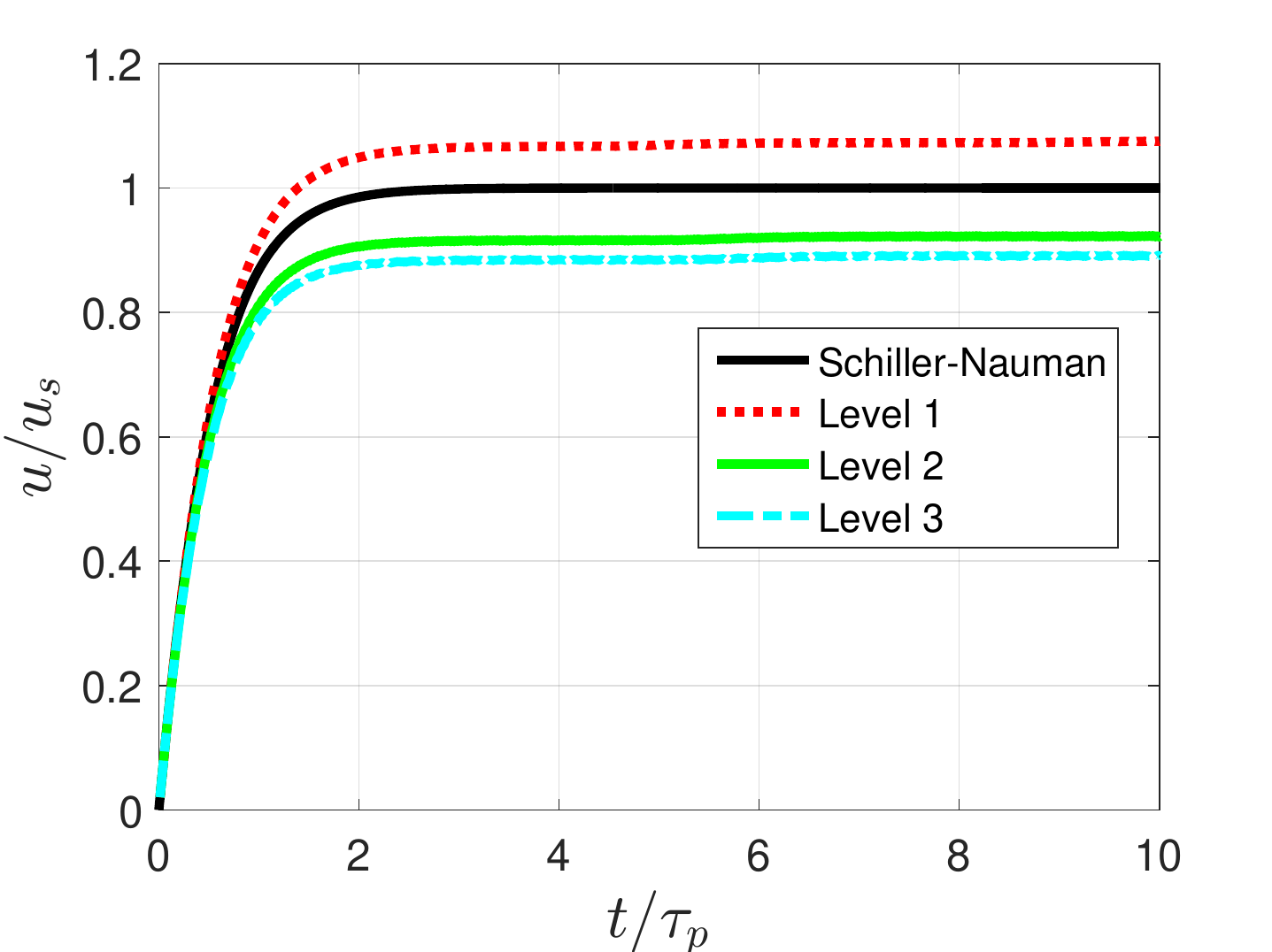} \label{fig:ref_L1_R10_St5}}
  \caption{Effect of $C$-field refinement on settling velocity histories, $\Lambda = 1$, $St_\Delta = 5$ (color online). $C$-field levels refer to coefficients given in Appendix B of \cite{horwitz-2016}.}
  \label{fig:fig4}
\end{figure}

Next, we examine the effect of $C$-field refinement on the settling velocity histories. These results are shown in Figure~\ref{fig:fig4}. For $Re_p = 0.1$, the settling velocity converges in steady-state to the analytical settling velocity. This observation is consistent with the streamlines being only slightly perturbed from the Stokesian streamlines when the Reynolds number is small. Alternatively, this is why $C$-field coefficients which assumed Stokesian symmetries were able to provide good predictions of the particle settling velocity even though a nonlinear drag correlation was used. In contrast however, at $Re_p = 10$, it is clear the correction developed for Stokes drag does not converge to the steady state settling velocity of the nonlinear drag correlation. This is a consequence of $C$-fields based on Stokesian symmetries  breaking down in the near-field of the particle. Since the $C$-field coefficients are tuned under the assumption of zero Reynolds number, convergence can only be expected in that limit. For precise prediction of the undisturbed fluid velocity outside the Stokes regime, it would be necessary to introduce Reynolds number dependent $C$-coefficients. Nevertheless, as we have shown, $C$-coefficients tuned in the zero Reynolds number limit yield reasonable results up to $Re_p = 10$. 


With regard to the applicability of two-way coupled point-particles in more complex applications, these observations suggest that the \enquote{contamination} effect, that is having numerically dependent features in the near field of point-particles, interacting with resolved features in the far field, may be of second order importance to accurate momentum/energy coupling. In \citet{mehrabadi-2016}, the authors find that point-particles obeying the Schiller-Nauman correlation calculated with the undisturbed fluid velocity reproduce well the TKE, dissipation, and particle kinetic energy, of that predicted in particle-resolved simulation of decaying particle-laden homogeneous isotropic turbulence. This suggests that two-way coupled point-particle simulations may have predictive power for integral statistics when the details of particle near fields are free from recirculation, below say, the formation of a wake region around $Re_p \approx 24$ \citep{taneda-1956}. One area that should be explored more is how the artificial structures created by point-particles affects more sensitive statistics like particle radial distribution function and Lagrangian structure functions. The main take away here is that the correction scheme compares reasonably well with the analytical solution at moderate Reynolds numbers (especially compared with the trilinear prediction in Figure $2$). 


\begin{figure}
  \centering
  \subfloat[$\Lambda = 0.5$] {\includegraphics[trim={3.5cm 8.5cm 3.5cm 8.5cm},clip, width=65mm]{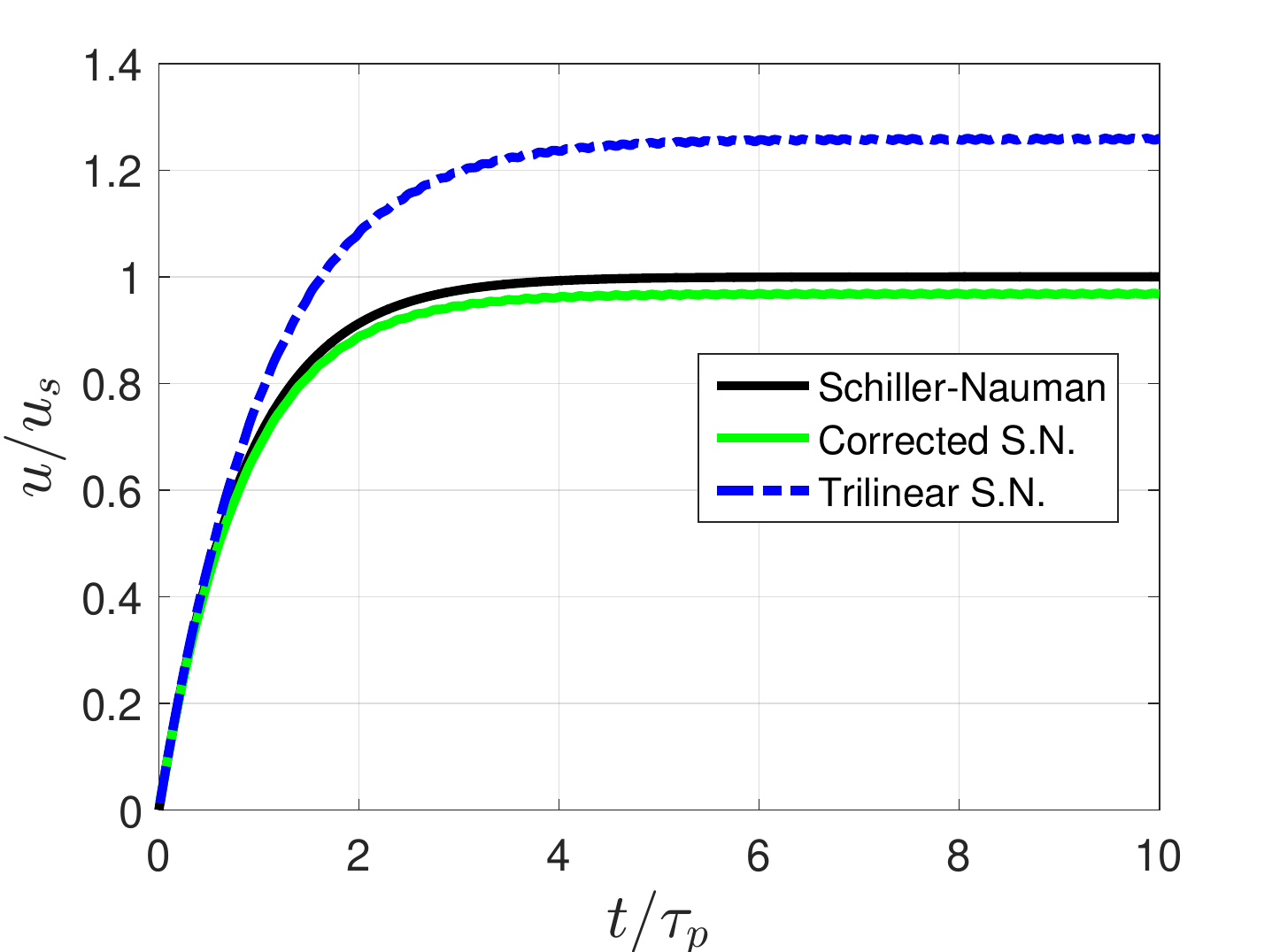} \label{fig:ref_L50_R1_St5}}
  \subfloat [$\Lambda = 0.1$] {\includegraphics[trim={3.5cm 8.5cm 3.5cm 8.5cm},clip, width=65mm]{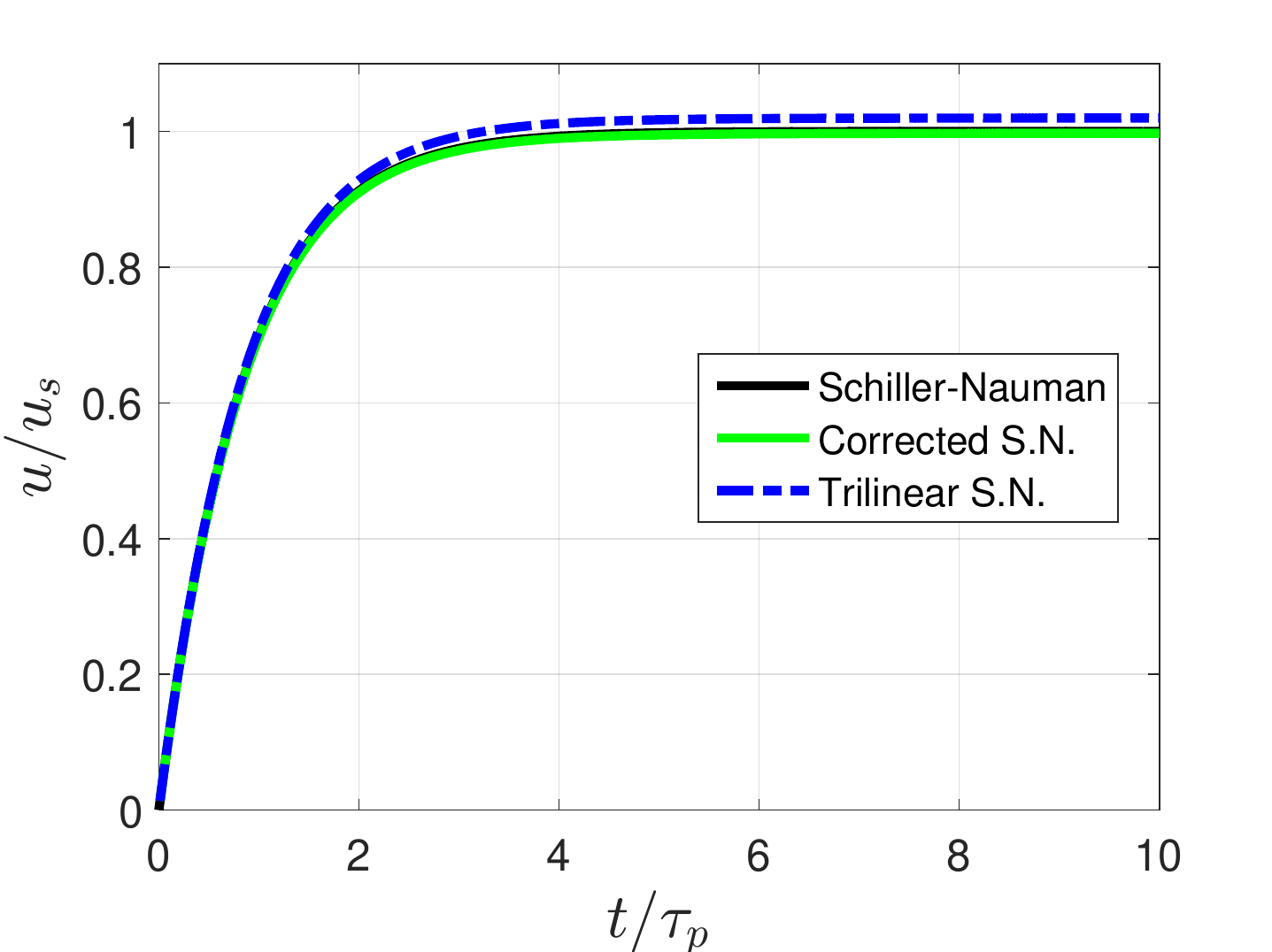} \label{fig:ref_L10_R1_St5}}
  \put(-135,25){\includegraphics[trim={4.2cm 9.2cm 4.2cm 9.2cm},clip, width=40mm]{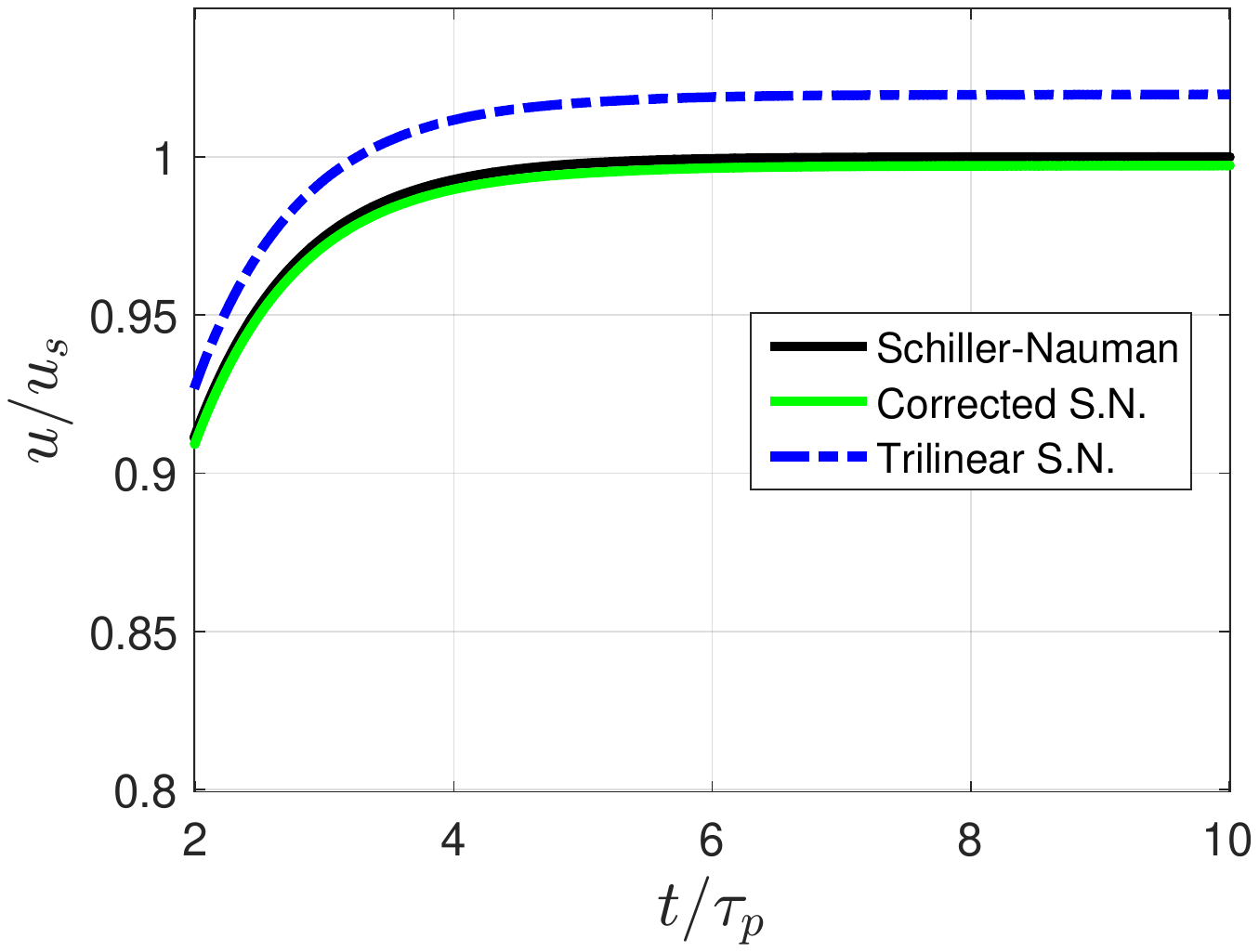}}
  \caption{Effect of non-dimensional particle size on settling velocity histories, $Re_p = 1$, $St_\Delta = 5$ (color online).}
  \label{fig:fig5}
\end{figure}

Figure~\ref{fig:fig5} demonstrates the robustness of the developed scheme in predicting the correct settling velocity over a range of dimensionless particle sizes  $\Lambda$. It is clear that the proposed correction performs better than trilinear interpolation at both $\Lambda = 0.5$ and $\Lambda = 0.1$. The steady state errors for the correction scheme are respectively $\approx 3\%$ and $0\%$ while for the trilinear scheme, the respective errors are $\approx 26\%$ and $2\%$.


\begin{figure}
  \centering
  \subfloat[$St_\Delta = 0.5, \Lambda = 1, \rho_p/\rho_f = 9$] {\includegraphics[trim={3.5cm 8.5cm 3.5cm 8.5cm},clip, width=65mm]{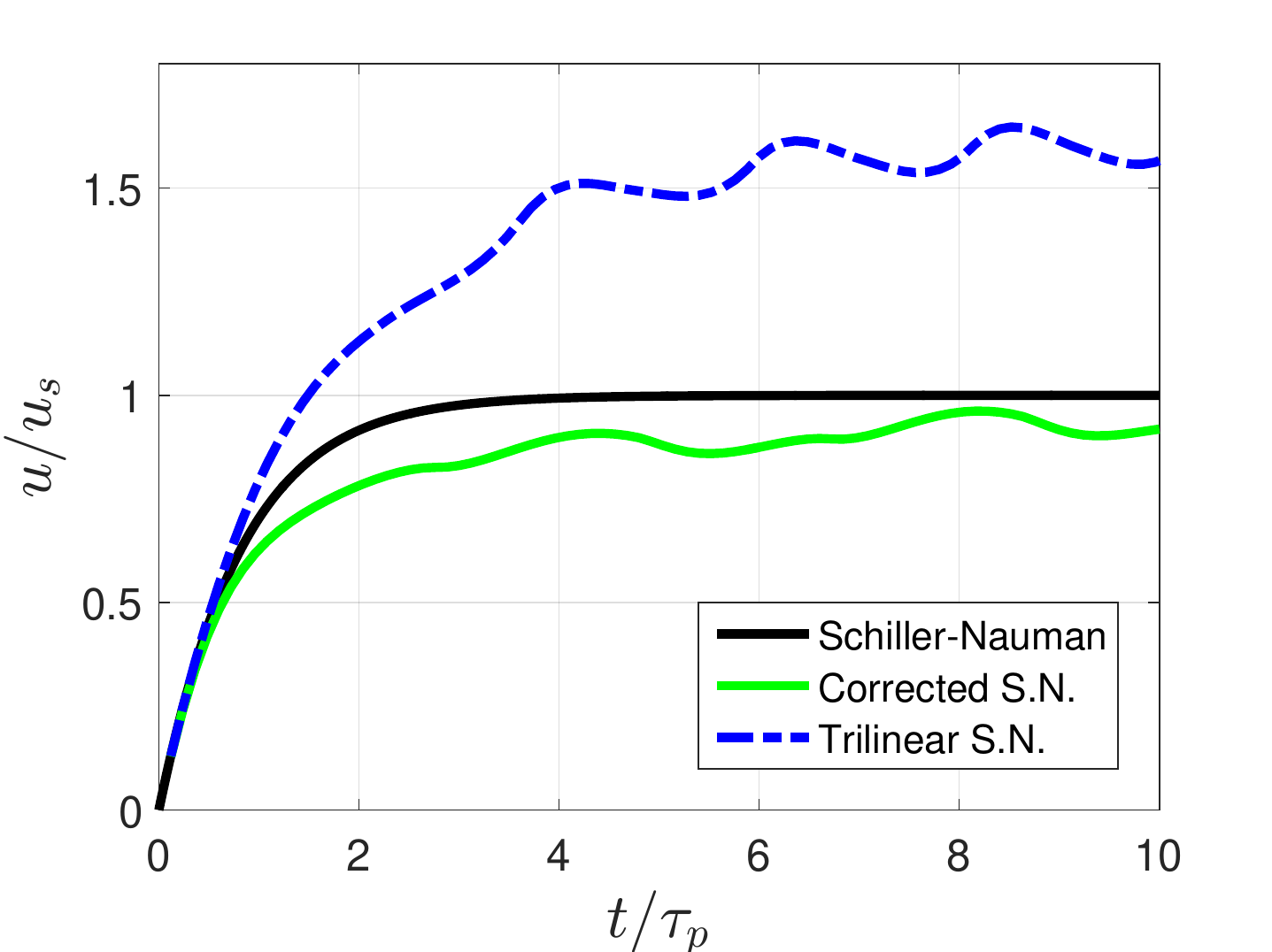} \label{fig:ref_L1_R1_St05}}
  \subfloat [$St_\Delta = 25, \Lambda = 1, \rho_p/\rho_f = 450$] {\includegraphics[trim={3.5cm 8.5cm 3.5cm 8.5cm},clip, width=65mm]{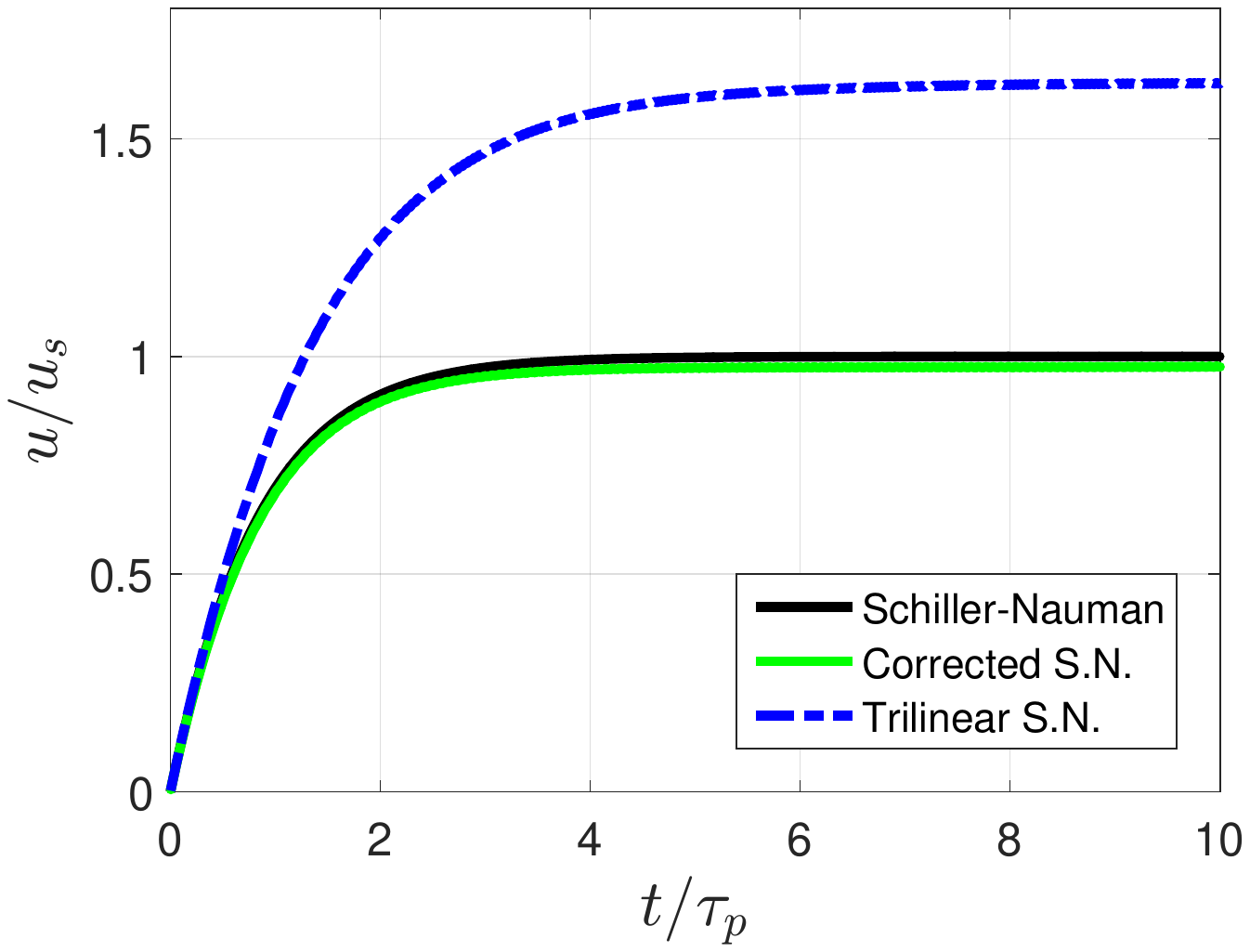} \label{fig:ref_L1_R1_St25}}

    \subfloat[$St_\Delta = 0.5, \Lambda = 0.5, \rho_p/\rho_f = 36$] {\includegraphics[trim={3.5cm 8.5cm 3.5cm 8.5cm},clip, width=65mm]{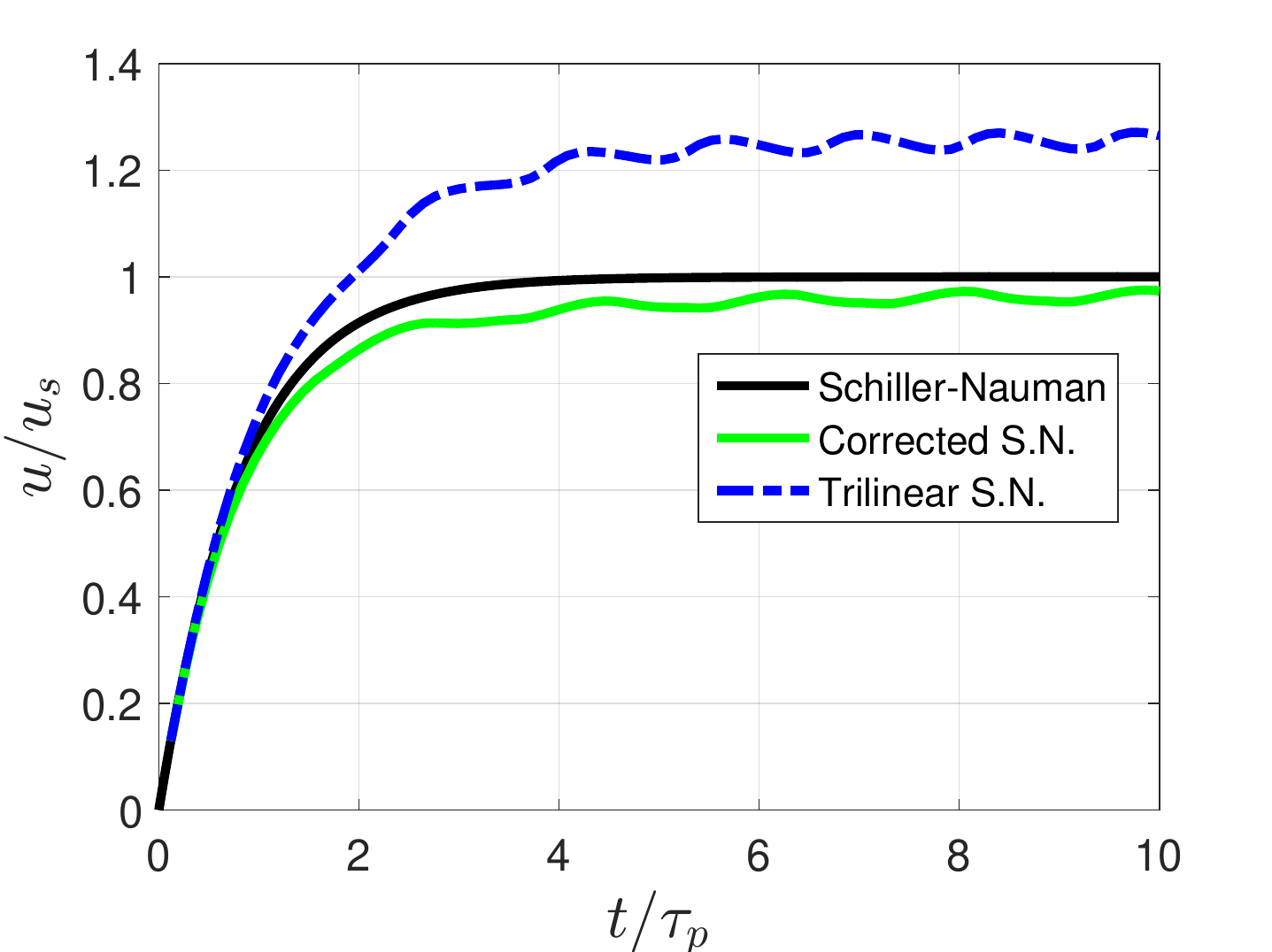} \label{fig:ref_L50_R1_St05}}
    \subfloat [$St_\Delta = 25, \Lambda = 0.5, \rho_p/\rho_f = 1800$] {\includegraphics[trim={3.5cm 8.5cm 3.5cm 8.5cm},clip, width=65mm]{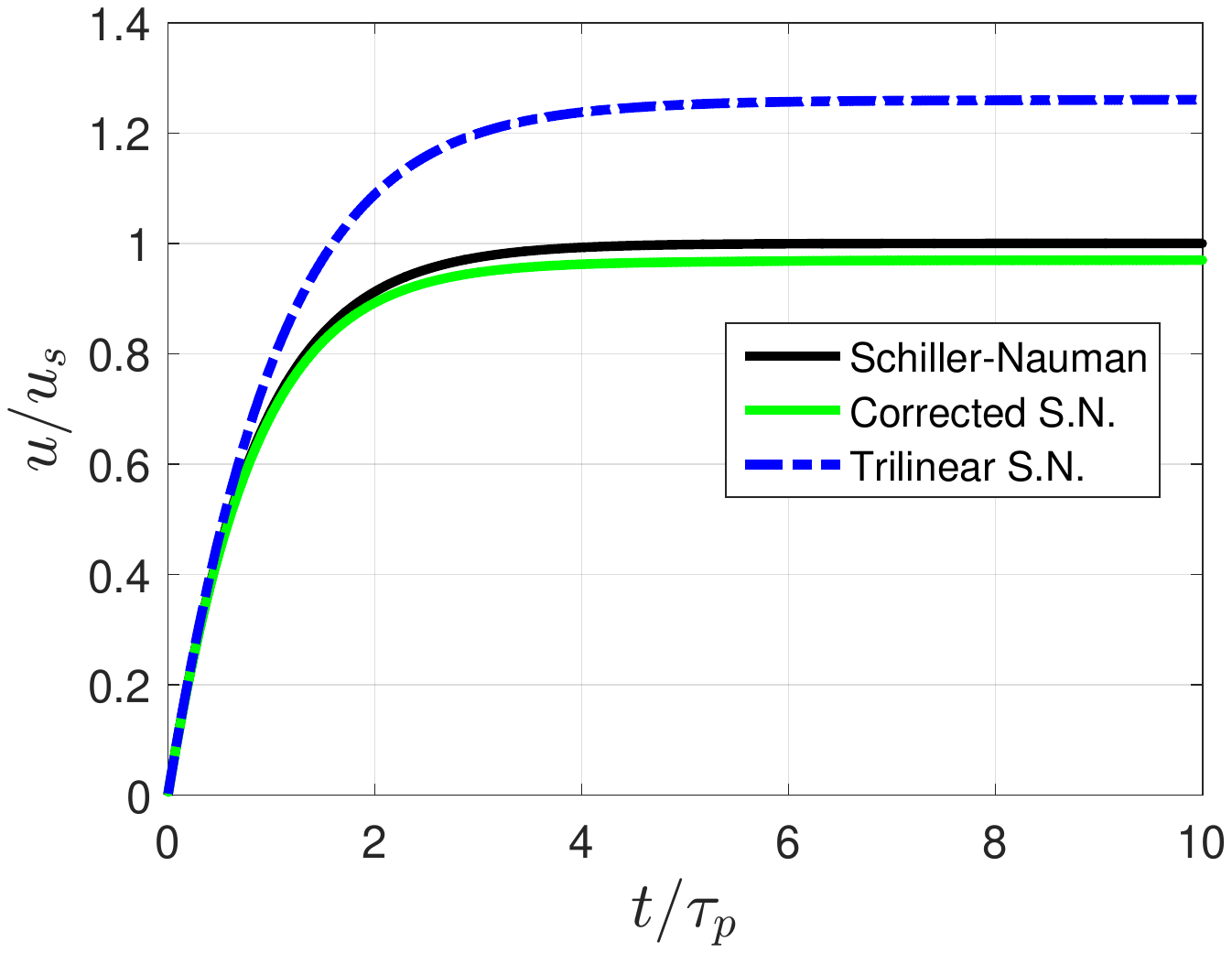} \label{fig:ref_L50_R1_St25}}
  \caption{Effect of Stokes number (density ratio) on settling velocity histories, $Re_p = 1$ (color online).}
  \label{fig:fig6}
\end{figure}

We also examine the effect of Stokes number on the settling velocity history. These results are shown in Figure~\ref{fig:fig6}. For both $St_\Delta = 0.5$ and $St_\Delta = 25$, the correction scheme compares much better with the analytical solution compared with the trilinear scheme. While the agreement for the high Stokes case is excellent, we see some undershoot in the settling velocity history predicted by the correction scheme for the low Stokes number case (Figure~\ref{fig:fig6} (a)). To simulate this particle with $\Lambda = 1$, it was necessary to reduce the particle to fluid density ratio to only 9. Therefore, the neglect of history effects \citep{Lovalenti-1993}, \citep{Maxey-1983}  over steady drag alone is not strictly justified, and the analytical solutions here have neglected history effects. While the undisturbed flow is identically zero (and therefore steady), the scheme used to estimate $\boldsymbol{\tilde{u}_p}$ is becoming more sensitive to numerical history effects owing to a particle seeing different parts of its own disturbance field within a grid cell, as the disturbance field is evolving in time. These observations were also observed in \citet{horwitz-2016}. We show in Figure~\ref{fig:fig6} (c) for a smaller non-dimensional grid size of $\Lambda = 0.5$, and density ratio of 36, the oscillations and undershoot in the steady state settling velocity are dramatically reduced. It appears the correction scheme is well-suited for the gas-solid flow regime where the density ratio is much larger than unity. However, even in cases well outside the gas-solid regime, e.g. $\rho_p/\rho_f = 36$, the correction scheme performs satisfactorily.
 In the limit of small Stokes number and high density ratio, ($St\ll1$) and ($\rho_p/\rho_f\gg1$), the non-dimensional particle size will inevitably be small ($\Lambda\ll1$),  so that the difference between the disturbed and undisturbed velocity will also be small.

\begin{figure}
   \centering
  \subfloat {\includegraphics[trim={3.5cm 8.0cm 3.5cm 8.0cm},clip, width=135mm]{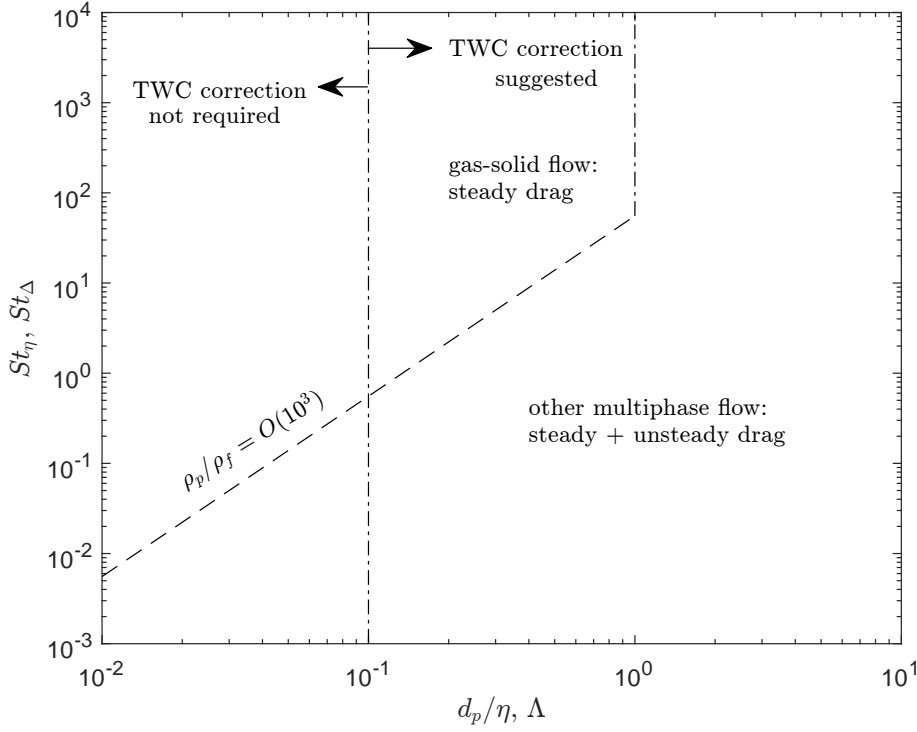} \label{fig:ref_regdiagram}}  
  \caption{Regime diagram for modelling point-particles.}
  \label{fig:fig7}
\end{figure}

The previous observations motivate a notional regime diagram for point-particle modelling (Figure~\ref{fig:fig7}). In \citet{horwitz-2016}, it was shown that the error in calculation of the undisturbed fluid velocity was $O(\Lambda)$. (Note the precise error will depend upon the projection stencil, the drag correlation, and the fluid numerics.) Using 10\% as an arbitrary boundary, the space of particle size and Stokes number can be divided into two regions: for $\Lambda < 0.1$, the undisturbed fluid velocity is comparable to the disturbed fluid velocity, so a correction to account for this difference may not be necessary. For $\Lambda > 0.1$, two-way coupling (TWC) effects are important in the near-field of the particle, so it is important that a correction for the undisturbed velocity be used, consistent with the drag model and projection scheme that are used. The parameter space may also be divided into two physical regimes based on density ratio. For $\rho_p/\rho_f \geq 1000$, the fluid-dispersed phase system is likely in the gas-solid regime. For smaller density ratios, the system is likely in another multiphase flow state such as gas-liquid, liquid-solid, liquid-liquid, or other more complicated mixture. In the gas-solid regime, steady drag (Stokes or Reynolds corrected) is likely to be  dominant over other drag terms. For example, non-dimensionalization of the Maxey-Riley equation reveals the fluid acceleration, added mass, and history terms are small at high density ratio. 

It is important to note that there is conflicting evidence as to the importance of the history term at high density ratio. \citet{Coimbra-1998} applied analytical Maxey-Riley solutions to particles moving in simple flows with and without the history term. At large density ratio, these two cases showed similar velocities, while \citet{Daitche-2015} explored particle-laden turbulent flow and showed, consistent with the analysis of \citet{Ling-2013}, that the critical parameter governing the importance of the history term was the particle size relative to the Kolmogorov scale. The respective results of \citet{Daitche-2015} regarding particle slip velocity and \citet{Olivieri-2014} concerning the radial distribution function of particles in turbulence with and without history effects show modest variation at a density ratio of $1000$, but significant variation at a density ratio of $10$. In view of the aforementioned scaling analysis, these observations may be confounded by the fact that the maximum particle diameter for the high density ratio particles was about one-tenth the Kolmogorov scale. Based on these observations, the regime diagram in Figure ~\ref{fig:fig7} has a cut off at a non-dimensional length-scale of unity, suggesting that, even in the case of gas-solid flow, unsteady drag may become significant owing to the particle size being comparable to or larger than characteristic fluid scales.

In contrast, in non-gas-solid flows, where the density of the dispersed phase is not much larger than that of the fluid phase, other drag terms are likely to be important. In such regimes, not only does the particle equation of motion require an undisturbed fluid velocity, but so too does it require, for example the undisturbed fluid acceleration. Proper implementation of this force requires a suitable correction scheme. As we have seen in the previous results, while the physical problem we have explored is the simple settling of a particle under gravity in an otherwise quiescent flow, the lack of undisturbed fluid time scales does not preclude numerical unsteadiness entering the solution. For the low Stokes number (low density ratio) settling cases, we saw undershooting as well as oscillations in the particle settling velocity. What our correction assumes, consistent with a high Stokes number or density ratio particle, is that a particle's disturbance flow has fully established, and then it begins to move. In other words, the disturbance field is always in equilibrium. This is what it means for a drag formula to be called ``steady''. When particle Stokes number is small or density ratio is low, now there is a coupled interaction since the particle begins to move before its disturbance flow has fully developed. Though the undisturbed flow has no interesting physics in this case, the numerically strange behavior is an alert that the physical assumptions provided for using the steady drag law were not justified at early times. So too then must modellers be aware that simply using a drag formula whose motivation is based on the undisturbed flow physics may fail to coupled interactions explicitly captured in the time history of the fluid phase (the $du/dt$ term is always solved for) and not explicitly captured in the particle model.



\section{Conclusion}
In this work, we tested a method used to predict the undisturbed fluid velocity created by Stokesian point-particles in a regime where the point-particles obey the Schiller-Nauman correlation. For different particle Reynolds numbers, non-dimensional sizes, and Stokes numbers, particles were released from rest in an otherwise stationary fluid and allowed to settle under gravity. When compared with trilinear interpolation to evaluate the fluid velocity found in the drag formula, the correction scheme offered significant improvement in predicting the particle settling velocity history for all of the parameters considered. The correction scheme was found to perform best in the high Stokes limit, but good agreement was also observed at low Stokes number with increasing particle to fluid density ratio. The correction scheme was found to predict the steady state settling velocity with reasonable accuracy ($<10 \%$), for Reynolds numbers up to ten. While the correction scheme \citep{horwitz-2016} was developed under the assumption of Stokesian symmetry, its ability to be used at finite particle Reynolds number in nonlinear correlations allows it to be used in a larger variety of applications.

We also propose a regime diagram to aid point-particle modellers. Its aim is to be used in conjunction with physical regime maps like that proposed by \cite{Elghobashi-2006}. Assuming that a modeller's problem falls in the two-way or four-way coupling regime, we suggest for what regimes the undisturbed fluid quantities require modelling, and which terms in the particle equation of motion should be included. In the case of four-way coupling, additional modelling concerns include collisions, Faxen, and lubrication effects owing to high volume fraction.

The type of correction scheme explored in this work also has relevance to multiphase problems in the presence of heat transfer. In these systems, the heat source would depend on the difference between the undisturbed fluid temperature and the particle's temperature. In this scenario, a correction scheme would be required to remove the temperature disturbance as seen by the particle that created it. Such a procedure would find applications in particle-laden solar receivers \citep{Pouransari-2017} where the particle concentration field can couple to solar radiation heating. A correction for the undisturbed temperature field would also be useful in predicting particle settling velocity in heated systems \citep{Frankel-2016} where particle heating of the gas can couple to dilitation modes with scale comparable to the particle size \citep{Pouransari-2017b}. 

\subsection*{Acknowledgments}
This work was funded by the United States Department of Energy through the 
Predictive Science Academic Alliance Program 2 (PSAAP2) at Stanford University. Jeremy Horwitz has also been supported by the National Science Foundation Graduate Research Fellowship under Grant No. DGE-114747. Any opinion, findings, and conclusions or recommendations expressed in this material are those of the authors and do not necessarily reflect the views of the National Science Foundation.


\bibliography{manuscript_draft}

\end{document}